\documentclass[paper,longauth,oneecolumn]{aa}
\usepackage[varg]{txfonts}
\usepackage[pdftex]{graphicx}
\usepackage{wasysym}
\usepackage[colorlinks]{hyperref}
\usepackage{url}
\usepackage{footmisc}
\usepackage{amssymb}
\usepackage{subfigure}
\usepackage{makecell}
\usepackage{multirow}
\usepackage{natbib}
\usepackage{soul, xcolor}
\RequirePackage[switch]{lineno}

\newcommand{\fermi}{$Fermi\mathrm{-LAT}$ }
\newcommand{\g}{$\gamma$}

\def\lsim{\mathrel{\rlap{\lower4pt\hbox{\hskip1pt$\sim$}}
    \raise1pt\hbox{$<$}}}                
\def\gsim{\mathrel{\rlap{\lower4pt\hbox{\hskip1pt$\sim$}}
    \raise1pt\hbox{$>$}}}                

\begin{document}
\setstcolor{red}
\title{The starburst galaxy NGC 253 revisited by H.E.S.S. and {\it Fermi}-LAT}

\titlerunning{Revisiting NGC\,253 with H.E.S.S.}
\authorrunning{H.E.S.S. Collaboration}

\author{H.E.S.S. Collaboration
\and H.~Abdalla \inst{\ref{NWU}}
\and F.~Aharonian \inst{\ref{MPIK},\ref{DIAS},\ref{NASRA}}
\and F.~Ait~Benkhali \inst{\ref{MPIK}}
\and E.O.~Ang\"uner \inst{\ref{CPPM}}
\and M.~Arakawa \inst{\ref{Rikkyo}}
\and C.~Arcaro \inst{\ref{NWU}}
\and C.~Armand \inst{\ref{LAPP}}
\and M.~Arrieta \inst{\ref{LUTH}}
\and M.~Backes \inst{\ref{UNAM},\ref{NWU}}
\and M.~Barnard \inst{\ref{NWU}}
\and Y.~Becherini \inst{\ref{Linnaeus}}
\and J.~Becker~Tjus \inst{\ref{RUB}}
\and D.~Berge \inst{\ref{DESY}}
\and S.~Bernhard \inst{\ref{LFUI}}
\and K.~Bernl\"ohr \inst{\ref{MPIK}}
\and R.~Blackwell \inst{\ref{Adelaide}}
\and M.~B\"ottcher \inst{\ref{NWU}}
\and C.~Boisson \inst{\ref{LUTH}}
\and J.~Bolmont \inst{\ref{LPNHE}}
\and S.~Bonnefoy \inst{\ref{DESY}}
\and P.~Bordas \inst{\ref{MPIK}}
\and J.~Bregeon \inst{\ref{LUPM}}
\and F.~Brun \inst{\ref{CENB}}
\and P.~Brun \inst{\ref{IRFU}}
\and M.~Bryan \inst{\ref{GRAPPA}}
\and M.~B\"{u}chele \inst{\ref{ECAP}}
\and T.~Bulik \inst{\ref{UWarsaw}}
\and T.~Bylund \inst{\ref{Linnaeus}}
\and M.~Capasso \inst{\ref{IAAT}}
\and S.~Caroff \inst{\ref{LLR}}
\and A.~Carosi \inst{\ref{LAPP}}
\and S.~Casanova \inst{\ref{IFJPAN},\ref{MPIK}}
\and M.~Cerruti \inst{\ref{LPNHE}}
\and N.~Chakraborty \inst{\ref{MPIK}}
\and S.~Chandra \inst{\ref{NWU}}
\and R.C.G.~Chaves \inst{\ref{LUPM},\ref{CurieChaves}}
\and A.~Chen \inst{\ref{WITS}}
\and S.~Colafrancesco \inst{\ref{WITS}}
\and B.~Condon \inst{\ref{CENB}}
\and I.D.~Davids \inst{\ref{UNAM}}
\and C.~Deil \inst{\ref{MPIK}}
\and J.~Devin \inst{\ref{LUPM}}
\and P.~deWilt \inst{\ref{Adelaide}}
\and L.~Dirson \inst{\ref{HH}}
\and A.~Djannati-Ata\"i \inst{\ref{APC}}
\and A.~Dmytriiev \inst{\ref{LUTH}}
\and A.~Donath \inst{\ref{MPIK}}
\and L.O'C.~Drury \inst{\ref{DIAS}}
\and J.~Dyks \inst{\ref{NCAC}}
\and K.~Egberts \inst{\ref{UP}}
\and G.~Emery \inst{\ref{LPNHE}}
\and J.-P.~Ernenwein \inst{\ref{CPPM}}
\and S.~Eschbach \inst{\ref{ECAP}}
\and S.~Fegan \inst{\ref{LLR}}
\and A.~Fiasson \inst{\ref{LAPP}}
\and G.~Fontaine \inst{\ref{LLR}}
\and S.~Funk \inst{\ref{ECAP}}
\and M.~F\"u{\ss}ling \inst{\ref{DESY}}
\and S.~Gabici \inst{\ref{APC}}
\and Y.A.~Gallant \inst{\ref{LUPM}}
\and T.~Garrigoux \inst{\ref{NWU}}
\and F.~Gat{\'e} \inst{\ref{LAPP}}
\and G.~Giavitto \inst{\ref{DESY}}
\and D.~Glawion \inst{\ref{LSW}}
\and J.F.~Glicenstein \inst{\ref{IRFU}}
\and D.~Gottschall \inst{\ref{IAAT}}
\and M.-H.~Grondin \inst{\ref{CENB}}
\and J.~Hahn \inst{\ref{MPIK}}
\and M.~Haupt \inst{\ref{DESY}}
\and G.~Heinzelmann \inst{\ref{HH}}
\and G.~Henri \inst{\ref{Grenoble}}
\and G.~Hermann \inst{\ref{MPIK}}
\and J.A.~Hinton \inst{\ref{MPIK}}
\and W.~Hofmann \inst{\ref{MPIK}}
\and C.~Hoischen \inst{\ref{UP}} \footnotemark[1]
\and T.~L.~Holch \inst{\ref{HUB}}
\and M.~Holler \inst{\ref{LFUI}}
\and D.~Horns \inst{\ref{HH}}
\and D.~Huber \inst{\ref{LFUI}}
\and H.~Iwasaki \inst{\ref{Rikkyo}}
\and A.~Jacholkowska \inst{\ref{LPNHE}} \protect\footnotemark[2] 
\and M.~Jamrozy \inst{\ref{UJK}}
\and D.~Jankowsky \inst{\ref{ECAP}}
\and F.~Jankowsky \inst{\ref{LSW}}
\and L.~Jouvin \inst{\ref{APC}}
\and I.~Jung-Richardt \inst{\ref{ECAP}}
\and M.A.~Kastendieck \inst{\ref{HH}}
\and K.~Katarzy{\'n}ski \inst{\ref{NCUT}}
\and M.~Katsuragawa \inst{\ref{JAXA}}
\and U.~Katz \inst{\ref{ECAP}}
\and D.~Kerszberg \inst{\ref{LPNHE}}
\and D.~Khangulyan \inst{\ref{Rikkyo}}
\and B.~Kh\'elifi \inst{\ref{APC}}
\and J.~King \inst{\ref{MPIK}}
\and S.~Klepser \inst{\ref{DESY}}
\and W.~Klu\'{z}niak \inst{\ref{NCAC}}
\and Nu.~Komin \inst{\ref{WITS}}
\and K.~Kosack \inst{\ref{IRFU}}
\and S.~Krakau \inst{\ref{RUB}}
\and M.~Kraus \inst{\ref{ECAP}}
\and P.P.~Kr\"uger \inst{\ref{NWU}}
\and G.~Lamanna \inst{\ref{LAPP}}
\and J.~Lau \inst{\ref{Adelaide}}
\and J.~Lefaucheur \inst{\ref{IRFU}}
\and A.~Lemi\`ere \inst{\ref{APC}}
\and M.~Lemoine-Goumard \inst{\ref{CENB}}
\and J.-P.~Lenain \inst{\ref{LPNHE}}
\and E.~Leser \inst{\ref{UP}}
\and T.~Lohse \inst{\ref{HUB}}
\and M.~Lorentz \inst{\ref{IRFU}}
\and R.~L\'opez-Coto \inst{\ref{MPIK}}
\and I.~Lypova \inst{\ref{DESY}}
\and D.~Malyshev \inst{\ref{IAAT}}
\and V.~Marandon \inst{\ref{MPIK}}
\and A.~Marcowith \inst{\ref{LUPM}}
\and C.~Mariaud \inst{\ref{LLR}}
\and G.~Mart\'i-Devesa \inst{\ref{LFUI}}
\and R.~Marx \inst{\ref{MPIK}}
\and G.~Maurin \inst{\ref{LAPP}}
\and P.J.~Meintjes \inst{\ref{UFS}}
\and A.M.W.~Mitchell \inst{\ref{MPIK}}
\and R.~Moderski \inst{\ref{NCAC}}
\and M.~Mohamed \inst{\ref{LSW}}
\and L.~Mohrmann \inst{\ref{ECAP}}
\and E.~Moulin \inst{\ref{IRFU}}
\and T.~Murach \inst{\ref{DESY}}
\and S.~Nakashima  \inst{\ref{JAXA}}
\and M.~de~Naurois \inst{\ref{LLR}}
\and H.~Ndiyavala  \inst{\ref{NWU}}
\and F.~Niederwanger \inst{\ref{LFUI}}
\and J.~Niemiec \inst{\ref{IFJPAN}}
\and L.~Oakes \inst{\ref{HUB}}
\and P.~O'Brien \inst{\ref{Leicester}}
\and H.~Odaka \inst{\ref{JAXA}}
\and S.~Ohm \inst{\ref{DESY}}\footnotemark[1]
\and M.~Ostrowski \inst{\ref{UJK}}
\and I.~Oya \inst{\ref{DESY}}
\and M.~Padovani \inst{\ref{LUPM}}
\and M.~Panter \inst{\ref{MPIK}}
\and R.D.~Parsons \inst{\ref{MPIK}}
\and C.~Perennes \inst{\ref{LPNHE}}
\and P.-O.~Petrucci \inst{\ref{Grenoble}}
\and B.~Peyaud \inst{\ref{IRFU}}
\and Q.~Piel \inst{\ref{LAPP}}
\and S.~Pita \inst{\ref{APC}}
\and V.~Poireau \inst{\ref{LAPP}}
\and A.~Priyana~Noel \inst{\ref{UJK}}
\and D.A.~Prokhorov \inst{\ref{WITS}}
\and H.~Prokoph \inst{\ref{DESY}}
\and G.~P\"uhlhofer \inst{\ref{IAAT}}
\and M.~Punch \inst{\ref{APC},\ref{Linnaeus}}
\and A.~Quirrenbach \inst{\ref{LSW}}
\and S.~Raab \inst{\ref{ECAP}}
\and R.~Rauth \inst{\ref{LFUI}}
\and A.~Reimer \inst{\ref{LFUI}}
\and O.~Reimer \inst{\ref{LFUI}}
\and M.~Renaud \inst{\ref{LUPM}}
\and F.~Rieger \inst{\ref{MPIK},\ref{FellowRieger}}
\and L.~Rinchiuso \inst{\ref{IRFU}}
\and C.~Romoli \inst{\ref{MPIK}}
\and G.~Rowell \inst{\ref{Adelaide}}
\and B.~Rudak \inst{\ref{NCAC}}
\and E.~Ruiz-Velasco \inst{\ref{MPIK}}
\and V.~Sahakian \inst{\ref{YPI},\ref{NASRA}}
\and S.~Saito \inst{\ref{Rikkyo}}
\and D.A.~Sanchez \inst{\ref{LAPP}}
\and A.~Santangelo \inst{\ref{IAAT}}
\and M.~Sasaki \inst{\ref{ECAP}}
\and R.~Schlickeiser \inst{\ref{RUB}}
\and F.~Sch\"ussler \inst{\ref{IRFU}}
\and A.~Schulz \inst{\ref{DESY}}
\and U.~Schwanke \inst{\ref{HUB}}
\and S.~Schwemmer \inst{\ref{LSW}}
\and M.~Seglar-Arroyo \inst{\ref{IRFU}}
\and M.~Senniappan \inst{\ref{Linnaeus}}
\and A.S.~Seyffert \inst{\ref{NWU}}
\and N.~Shafi \inst{\ref{WITS}}
\and I.~Shilon \inst{\ref{ECAP}}
\and K.~Shiningayamwe \inst{\ref{UNAM}}
\and R.~Simoni \inst{\ref{GRAPPA}}
\and A.~Sinha \inst{\ref{APC}}
\and H.~Sol \inst{\ref{LUTH}}
\and F.~Spanier \inst{\ref{NWU}}
\and A.~Specovius \inst{\ref{ECAP}}
\and M.~Spir-Jacob \inst{\ref{APC}}
\and {\L.}~Stawarz \inst{\ref{UJK}}
\and R.~Steenkamp \inst{\ref{UNAM}}
\and C.~Stegmann \inst{\ref{UP},\ref{DESY}}
\and C.~Steppa \inst{\ref{UP}}
\and I.~Sushch \inst{\ref{NWU}}
\and T.~Takahashi  \inst{\ref{JAXA}}
\and J.-P.~Tavernet \inst{\ref{LPNHE}}
\and T.~Tavernier \inst{\ref{IRFU}}
\and A.M.~Taylor \inst{\ref{DESY}}\footnotemark[1]
\and R.~Terrier \inst{\ref{APC}}
\and L.~Tibaldo \inst{\ref{MPIK}}
\and D.~Tiziani \inst{\ref{ECAP}}
\and M.~Tluczykont \inst{\ref{HH}}
\and C.~Trichard \inst{\ref{LLR}}
\and M.~Tsirou \inst{\ref{LUPM}}
\and N.~Tsuji \inst{\ref{Rikkyo}}
\and R.~Tuffs \inst{\ref{MPIK}}
\and Y.~Uchiyama \inst{\ref{Rikkyo}}
\and D.J.~van~der~Walt \inst{\ref{NWU}}
\and C.~van~Eldik \inst{\ref{ECAP}}
\and C.~van~Rensburg \inst{\ref{NWU}}
\and B.~van~Soelen \inst{\ref{UFS}}
\and G.~Vasileiadis \inst{\ref{LUPM}}
\and J.~Veh \inst{\ref{ECAP}}
\and C.~Venter \inst{\ref{NWU}}
\and A.~Viana \inst{\ref{MPIK},\ref{VianaNowAt}}
\and P.~Vincent \inst{\ref{LPNHE}}
\and J.~Vink \inst{\ref{GRAPPA}}
\and F.~Voisin \inst{\ref{Adelaide}}
\and H.J.~V\"olk \inst{\ref{MPIK}}\footnotemark[1]
\and T.~Vuillaume \inst{\ref{LAPP}}
\and Z.~Wadiasingh \inst{\ref{NWU}}
\and S.J.~Wagner \inst{\ref{LSW}}
\and P.~Wagner \inst{\ref{HUB}}
\and R.M.~Wagner \inst{\ref{OKC}}
\and R.~White \inst{\ref{MPIK}}
\and A.~Wierzcholska \inst{\ref{IFJPAN}}
\and A.~W\"ornlein \inst{\ref{ECAP}}
\and R.~Yang \inst{\ref{MPIK}}\footnotemark[1]
\and D.~Zaborov \inst{\ref{LLR}}
\and M.~Zacharias \inst{\ref{NWU}}
\and R.~Zanin \inst{\ref{MPIK}}
\and A.A.~Zdziarski \inst{\ref{NCAC}}
\and A.~Zech \inst{\ref{LUTH}}
\and F.~Zefi \inst{\ref{LLR}}
\and A.~Ziegler \inst{\ref{ECAP}}
\and J.~Zorn \inst{\ref{MPIK}}
\and N.~\.Zywucka \inst{\ref{UJK}}
}

\institute{
Centre for Space Research, North-West University, Potchefstroom 2520, South Africa \label{NWU} \and 
Universit\"at Hamburg, Institut f\"ur Experimentalphysik, Luruper Chaussee 149, D 22761 Hamburg, Germany \label{HH} \and 
Max-Planck-Institut f\"ur Kernphysik, P.O. Box 103980, D 69029 Heidelberg, Germany \label{MPIK} \and 
Dublin Institute for Advanced Studies, 31 Fitzwilliam Place, Dublin 2, Ireland \label{DIAS} \and 
National Academy of Sciences of the Republic of Armenia,  Marshall Baghramian Avenue, 24, 0019 Yerevan, Republic of Armenia  \label{NASRA} \and
Yerevan Physics Institute, 2 Alikhanian Brothers St., 375036 Yerevan, Armenia \label{YPI} \and
Institut f\"ur Physik, Humboldt-Universit\"at zu Berlin, Newtonstr. 15, D 12489 Berlin, Germany \label{HUB} \and
University of Namibia, Department of Physics, Private Bag 13301, Windhoek, Namibia \label{UNAM} \and
GRAPPA, Anton Pannekoek Institute for Astronomy, University of Amsterdam,  Science Park 904, 1098 XH Amsterdam, The Netherlands \label{GRAPPA} \and
Department of Physics and Electrical Engineering, Linnaeus University,  351 95 V\"axj\"o, Sweden \label{Linnaeus} \and
Institut f\"ur Theoretische Physik, Lehrstuhl IV: Weltraum und Astrophysik, Ruhr-Universit\"at Bochum, D 44780 Bochum, Germany \label{RUB} \and
Institut f\"ur Astro- und Teilchenphysik, Leopold-Franzens-Universit\"at Innsbruck, A-6020 Innsbruck, Austria \label{LFUI} \and
School of Physical Sciences, University of Adelaide, Adelaide 5005, Australia \label{Adelaide} \and
LUTH, Observatoire de Paris, PSL Research University, CNRS, Universit\'e Paris Diderot, 5 Place Jules Janssen, 92190 Meudon, France \label{LUTH} \and
Sorbonne Universit\'e, Universit\'e Paris Diderot, Sorbonne Paris Cit\'e, CNRS/IN2P3, Laboratoire de Physique Nucl\'eaire et de Hautes Energies, LPNHE, 4 Place Jussieu, F-75252 Paris, France \label{LPNHE} \and
Laboratoire Univers et Particules de Montpellier, Universit\'e Montpellier, CNRS/IN2P3,  CC 72, Place Eug\`ene Bataillon, F-34095 Montpellier Cedex 5, France \label{LUPM} \and
IRFU, CEA, Universit\'e Paris-Saclay, F-91191 Gif-sur-Yvette, France \label{IRFU} \and
Astronomical Observatory, The University of Warsaw, Al. Ujazdowskie 4, 00-478 Warsaw, Poland \label{UWarsaw} \and
Aix Marseille Universit\'e, CNRS/IN2P3, CPPM, Marseille, France \label{CPPM} \and
Instytut Fizyki J\c{a}drowej PAN, ul. Radzikowskiego 152, 31-342 Krak{\'o}w, Poland \label{IFJPAN} \and
Funded by EU FP7 Marie Curie, grant agreement No. PIEF-GA-2012-332350 \label{CurieChaves}  \and
School of Physics, University of the Witwatersrand, 1 Jan Smuts Avenue, Braamfontein, Johannesburg, 2050 South Africa \label{WITS} \and
Laboratoire d'Annecy de Physique des Particules, Univ. Grenoble Alpes, Univ. Savoie Mont Blanc, CNRS, LAPP, 74000 Annecy, France \label{LAPP} \and
Landessternwarte, Universit\"at Heidelberg, K\"onigstuhl, D 69117 Heidelberg, Germany \label{LSW} \and
Universit\'e Bordeaux, CNRS/IN2P3, Centre d'\'Etudes Nucl\'eaires de Bordeaux Gradignan, 33175 Gradignan, France \label{CENB} \and
Oskar Klein Centre, Department of Physics, Stockholm University, Albanova University Center, SE-10691 Stockholm, Sweden \label{OKC} \and
Institut f\"ur Astronomie und Astrophysik, Universit\"at T\"ubingen, Sand 1, D 72076 T\"ubingen, Germany \label{IAAT} \and
Laboratoire Leprince-Ringuet, Ecole Polytechnique, CNRS/IN2P3, F-91128 Palaiseau, France \label{LLR} \and
APC, AstroParticule et Cosmologie, Universit\'{e} Paris Diderot, CNRS/IN2P3, CEA/Irfu, Observatoire de Paris, Sorbonne Paris Cit\'{e}, 10, rue Alice Domon et L\'{e}onie Duquet, 75205 Paris Cedex 13, France \label{APC} \and
Univ. Grenoble Alpes, CNRS, IPAG, F-38000 Grenoble, France \label{Grenoble} \and
Department of Physics and Astronomy, The University of Leicester, University Road, Leicester, LE1 7RH, United Kingdom \label{Leicester} \and
Nicolaus Copernicus Astronomical Center, Polish Academy of Sciences, ul. Bartycka 18, 00-716 Warsaw, Poland \label{NCAC} \and
Institut f\"ur Physik und Astronomie, Universit\"at Potsdam,  Karl-Liebknecht-Strasse 24/25, D 14476 Potsdam, Germany \label{UP} \and
Friedrich-Alexander-Universit\"at Erlangen-N\"urnberg, Erlangen Centre for Astroparticle Physics, Erwin-Rommel-Str. 1, D 91058 Erlangen, Germany \label{ECAP} \and
DESY, D-15738 Zeuthen, Germany \label{DESY} \and
Obserwatorium Astronomiczne, Uniwersytet Jagiello{\'n}ski, ul. Orla 171, 30-244 Krak{\'o}w, Poland \label{UJK} \and
Centre for Astronomy, Faculty of Physics, Astronomy and Informatics, Nicolaus Copernicus University,  Grudziadzka 5, 87-100 Torun, Poland \label{NCUT} \and
Department of Physics, University of the Free State,  PO Box 339, Bloemfontein 9300, South Africa \label{UFS} \and
Heisenberg Fellow (DFG), ITA Universit\"at Heidelberg, Germany \label{FellowRieger} \and
Department of Physics, Rikkyo University, 3-34-1 Nishi-Ikebukuro, Toshima-ku, Tokyo 171-8501, Japan \label{Rikkyo} \and
Japan Aerospace Exploration Agency (JAXA), Institute of Space and Astronautical Science (ISAS), 3-1-1 Yoshinodai, Chuo-ku, Sagamihara, Kanagawa 229-8510,  Japan \label{JAXA} \and
Now at The School of Physics, The University of New South Wales, Sydney, 2052, Australia \label{MaxtedNowAt} \and
Now at Instituto de F\'{i}sica de S\~{a}o Carlos, Universidade de S\~{a}o Paulo, Av. Trabalhador S\~{a}o-carlense, 400 - CEP 13566-590, S\~{a}o Carlos, SP, Brazil \label{VianaNowAt} 
}

\offprints{H.E.S.S.~collaboration,
\protect\\\email{\href{mailto:contact.hess@hess-experiment.eu}{contact.hess@hess-experiment.eu}};
\protect\\\protect\footnotemark[1] Corresponding authors
\protect\\\protect\footnotemark[2] Deceased
}
\date{}
\abstract
{NGC\,253 is one of only two starburst galaxies found to emit
  \g-rays from hundreds of MeV to multi-TeV energies. Accurate
  measurements of the very-high-energy (VHE) (E $>$ 100 GeV)
  and high-energy (HE) (E $>$ 60 MeV) spectra are crucial to
  study the underlying particle accelerators, probe the dominant
  emission mechanism(s) and to study cosmic-ray interaction and
  transport.}
{The measurement of the VHE \g-ray emission of NGC\,253 published in
  2012 by H.E.S.S. was limited by large systematic uncertainties.
  Here, the most up to date measurement of the \g-ray spectrum of
  NGC\,253 is investigated in both HE and VHE \g-rays. Assuming a
  hadronic origin of the \g-ray emission, the measurement
  uncertainties are propagated into the interpretation of the
  accelerated particle population.}
{The data of H.E.S.S. observations are reanalysed using an updated
  calibration and analysis chain. The improved \fermi analysis employs
  more than 8 years of data processed using pass 8. The cosmic-ray
  particle population is evaluated from the combined HE--VHE \g-ray
  spectrum using NAIMA in the optically thin case.}
{The VHE \g-ray energy spectrum is best fit by a power-law
  distribution with a flux normalisation of
  \mbox{$(1.34\,\pm\,0.14^{\mathrm{stat}}\,\pm\,0.27^{\mathrm{sys}}) \times
  10^{-13} \mathrm{cm^{-2} s^{-1} TeV^{-1}}$} at 1\,TeV --
  about 40\,\% above, but compatible with the value obtained in
  Abramowski et al. (2012). The spectral index
  \mbox{$\Gamma = 2.39 \pm 0.14^{\mathrm{stat}} \pm 0.25^{\mathrm{sys}}$} is
  slightly softer than but consistent with the previous measurement within
  systematic errors. In the Fermi energy range an integral flux of
  \mbox{$F(E>60\,{\rm MeV}) =
  (1.56\pm0.28^{\mathrm{stat}}\pm0.15^{\mathrm{sys}}) \times
  10^{-8}\,{\rm cm^{-2} s^{-1}}$} is obtained. At energies above $\sim$\,3 GeV the
  HE spectrum is consistent with a power-law ranging into the VHE part
  of the spectrum measured by H.E.S.S. with an overall spectral index
  $\Gamma = 2.22 \pm 0.06^{\rm stat}$}
{ Two scenarios for the starburst nucleus are tested, in which the gas
  in the starburst nucleus acts as either a thin or a thick target for
  hadronic cosmic rays accelerated by the individual sources in the nucleus.
  In these two models, the level to which NGC\,253 acts as a
  calorimeter is estimated to a range of $f_{\rm cal} = 0.1$ to $1$
  while accounting for the measurement uncertainties. The presented
  spectrum is likely to remain the most accurate measurements until
  the Cherenkov Telescope Array (CTA) has collected a substantial
  set of data towards NGC\,253.}
\maketitle
  

\section{Introduction}

Starburst galaxies are characterised by their high star-formation
rate (SFR) and gas-consumption times of 1\,Gyr or less. The starburst
phase typically lasts for a few hundred million years \citep[see e.g.][and references
therein]{Kennicutt2012, Krumholz2014}. Supernova (SN) remnants are
believed to be the main sources of the Galactic cosmic rays (CRs).
Starburst galaxies with their enhanced SFR and SN rate provide a
testbed to probe this paradigm. Furthermore, CRs are star-formation
regulators and drive complex chemical reactions by penetrating deep
into dense molecular cloud cores \citep[e.g.][]{Indriolo2013}. There
is also increasing evidence that CRs play an important role in galaxy
formation and evolution \citep{Booth2013, Salem2014a, salem2016role, Pakmor2016} by
driving galactic winds along expanding magnetic loops
\citep[e.g.][]{breitschwerdt1991galactic, breitschwerdt1993galactic}
that result from their excitation of a Parker instability in the disk
\citep{parker1966dynamical}. In this process the CRs heat the
outflowing gas through the non-linear Landau damping of the scattering
Alfv${\rm \acute{e}}$n waves that are excited by the outward streaming
CRs \citep[e.g.][]{zirakashvili1996magnetohydrodynamic}. This CR
heating might even prevent low-mass star formation in regions of very
high CR densities such as starburst galaxies
\citep[e.g.][]{Papadopoulos2013}.

Observations of starburst galaxies at \g-ray energies provide a useful
probe to test the physics of CRs in these objects: i) They permit
inference on the efficiency with which kinetic energy released in SN
explosions is channelled via relativistic particles into \g-rays. ii)
\g-rays can be used to infer properties of the interstellar medium in
starburst galaxies or probe energy partition between CRs,
magnetic fields and radiation fields. iii) Finally, \g-ray
measurements can be used as an independent probe to test the paradigm
of CR acceleration in SN remnant shocks.

The two archetypical starburst galaxies NGC\,253 and M82 have been
discovered to emit \g-rays with energies ranging from hundreds of MeV
to several TeV \citep{acero2009detection, acciari2009,
  abdo2010}. Subsequently, detailed spectral studies of NGC\,253 at
TeV energies \citep[][HESS12 in the following]{abramowski2012spectral}
and the systematic search for GeV \g-ray emission from a sample of
star-forming galaxies with \fermi, including NGC\,253 and
M82~\citep{ackermann2012gev}, have been presented. Recently, NGC\,253
has also been studied at hard X-rays with {\it NuSTAR}, soft X-rays
with {\it Chandra} and at radio wavelengths with {\it VLBA}
\citep{Wik2014}. Broadband spectral energy distribution (SED)
modelling is performed with different approaches, ranging from
semi-analytical one-zone models as described in
e.g. \citet{Eichmann2016}, to three-dimensional (3D) steady state models
\citep[e.g.][]{Persic2008, Rephaeli2010} and the treatment of time and
space-dependent injectors \citep[e.g.][]{Torres2012}. Starburst
galaxies are also discussed as one of the possible source classes
contributing to the astrophysical neutrino excess seen by the IceCube
collaboration\,\citep{Aartsen2014}. NGC\,253 remains one of the
weakest detected TeV \g-ray sources. After three years of improvements
to the calibration, reconstruction and analysis, a re-analysis of the
\g-ray data, including a re-evaluation of systematic uncertainties, is
presented in this work. In addition, 8 years of \fermi data are
analysed and the updated \g-ray spectrum from 60\,MeV to more than
10\,TeV is studied. As a result of both improvements, the discussion
presented in HESS12 is revisited.


\section{H.E.S.S. data analysis}
\label{sec:hess}

\subsection{H.E.S.S. data}

H.E.S.S. is an array of imaging atmospheric Cherenkov telescopes
located in the Khomas Highland of Namibia and detects cosmic \g-rays
with energies ranging from $\sim$0.1 to $\sim$100\,TeV. The data used
in this work are identical to the data presented and analysed in
HESS12. The observations carried out in 2005 and from 2007 to 2009 sum
to 158\,h of acceptance-corrected live time.  NGC\,253 has not been
the target of new observations from H.E.S.S. since then. Significantly improved
statistics would only be possible at the cost of unreasonably large amounts of
observation time. For a detailed description of the instrument and data set, 
the reader is referred to the original publication. The differences and
improvements of the analysis methods compared to HESS12 are highlighted
where applicable.

\subsection{H.E.S.S. analysis}
\label{sec:hess_analysis}
The results presented here and in HESS12 are based on a
semi-analytical model of air showers for the event reconstruction and
background suppression \citep{de2009high}. This {\it model analysis}
provides an improved angular resolution and a much better sensitivity
compared to the classical Hillas parameter-based analysis. It is,
however, susceptible to imperfections in the detailed modelling of the
instrument response. Since the original publication, a small
misalignment of the camera's position with respect to the telescope
dish was found. This was not fully taken into account in the HESS12
analysis but has been accounted for in newer versions of the analysis
software. The improvements in the pointing model have been verified
using optical star positions and by application to known, strong
\g-ray sources. The misaligned cameras resulted in a broadening of the
point-spread function (PSF) and introduced a shift of the main
discrimination variable in the {\it model analysis}. This shift led to
misclassifications of \g-rays as background and resulted in an
underestimation of the \g-ray flux of NGC\,253. The same behaviour was
uncovered and accounted for during the analysis of N\,157B in the
Large Magellanic Cloud \citep[supplement]{LMC2015}. The cross-check
analysis presented in HESS12 is not significantly affected by the
imperfect pointing model. The resulting systematic difference in
reconstructed \g-ray flux between the {\it model analysis} and the
cross-check analysis was taken into account in the flux systematic
uncertainty in HESS12. As the precision of the \g-ray spectrum
presented in HESS12 was limited by the systematic flux uncertainty,
and since the modelling of the camera positions were improved since
then, a reanalysis of the NGC\,253 H.E.S.S. data and revised VHE
\g-ray spectrum using the same data set as used in HESS12 is
presented.

\subsubsection{VHE \g-ray spectrum}
The data reanalysis was performed using the same analysis framework as
in HESS12, namely the {\it model analysis} \citep{de2009high}. An
updated position, extension limit, light curve and \g-ray spectrum are
derived. The source is detected with a slightly lower significance of
7.2\,$\sigma$ compared to $8.4\,\sigma$ in HESS12. The updated source
position is
$\mathrm{RA} = 0\mathrm{h}\,47\mathrm{m}\,32.54\mathrm{s} \pm
0\mathrm{m}\,11.2\mathrm{s}$,
$\mathrm{Dec} = -25\mathrm{d}\,17'\,25.4'' \pm 0'\,10.3''$ (J2000),
which changed only marginally towards an even better agreement with
the optical center of NGC\,253 at
$\mathrm{RA} = 0\mathrm{h}\,47\mathrm{m}\,33.1\mathrm{s}$,
$\mathrm{Dec} = -25\mathrm{d}\,17'\,18''$. With the PSF being
understood better, a possible extension of the \g-ray source is
constrained to $\leq1.4'$ at the $3\,\sigma$ level compared to the
previous limit of $\leq2.4'$. The new \g-ray spectrum, extracted at
the best-fit position, is well described by a single power law,
depicted in Fig.~\ref{fig:combinedFit}. The spectral points are
given in Table~\ref{tab:spec_points}. The flux normalisation
\mbox{$F_0 (1\,\mathrm{TeV}) =
  (1.34\,\pm\,0.14^{\mathrm{stat}}\,\pm\,0.27^{\mathrm{sys}}) \times
  10^{-13}\,\mathrm{cm^{-2}\,s^{-1}\,TeV^{-1}}$} is $\sim$\,40\,\%
higher, and the best-fit spectrum is with a spectral index of
$\Gamma=2.39 \pm 0.14^{\mathrm{stat}} \pm 0.25^{\mathrm{sys}}$
somewhat softer but consistent within errors compared to HESS12, where
a spectral index of
$\Gamma^{2012} = 2.14 \, \pm 0.18^{\rm stat} \, \pm 0.30^{\rm sys}$
and a normalisation at 1 TeV of
$F_{0}^{2012} = (9.6 \pm 1.5^{\rm stat}\, (+ 5.7, -2.9)^{\rm sys})
\times 10^{-14}\,{\rm TeV^{-1}\,cm^{-2}\,s^{-1}}$ were reported. Both
spectral parameters are consistent within the previously estimated
systematic uncertainties. The relative statistical errors are slightly
reduced due to the higher reconstructed \g-ray flux. We note that the
systematic uncertainties are now comparable to the statistical
uncertainties.

\subsubsection{Estimation of systematic uncertainties}
\label{sec:hess_systematics}
In HESS12, systematic uncertainties were estimated using a cross-check
analysis, which accounted for systematic differences of the
calibration, reconstruction and background subtraction. As this
cross-check analysis proved to be unaffected by the imperfections in
the modelling of the camera positions described above, the updated
analysis presented here is compared to the original cross-check
analysis. In HESS12, the difference in the flux normalisation between
the two analysis chains was found to be $50\,\%$, while the
re-analysed flux normalisations agree within 2\,\%. The difference of
best-fit spectral indices is on the 10\,\% level. As a 2\,\%
difference is likely not representative for the real systematic
uncertainty caused by different calibration chains, event
reconstruction and background subtraction procedures, additional tests
for systematic effects within the primary analysis framework have been
performed.

A test for systematic shifts in the reconstructed \g-ray flux caused
by the run selection has been performed by applying the original data
quality criteria \citep[e.g.][]{crabpaper} used in HESS12 in
comparison to a revised data quality selection. We found the data
selection has an impact on the reconstructed flux at a level of 10\,\% and
3\,\% in flux normalisation and spectral index, respectively,
for this faint source.

The applied \g-ray selection cuts might also introduce systematic
effects. To estimate the impact of the chosen cuts, the data set was
analysed using two different cut configurations: one designed for a
low-energy threshold and optimised for spectral studies
\textit{(Standard)}, the other optimised for weak sources (like
NGC\,253) and morphological studies, with a higher energy threshold
\textit{(Faint)}. The differences between the analyses with the two
cut configurations are 5\,\% in the spectral index and 13\,\% in the
normalisation, and represent an estimate of the systematic uncertainty
associated to the specific choice of the cut configuration.

The atmosphere is an integral part of an imaging atmospheric Cherenkov
telescope and varies over time. The assumed atmospheric density
profile influences the amount of light predicted to be seen by each
camera. The light yield is uncertain by $\sim$10\,\%
\citep{crabpaper}. In order to estimate the effect of this uncertainty
on the spectral parameters, the fit was repeated using response
functions that were shifted by $\pm\,10\,\%$ in energy. The resulting
uncertainties are 10\,\% and $\pm\,0.09$ for the flux normalisation
and the spectral index, respectively.

\begin{figure*}[tb]
\begin{center}
\resizebox{\hsize}{!}{\includegraphics{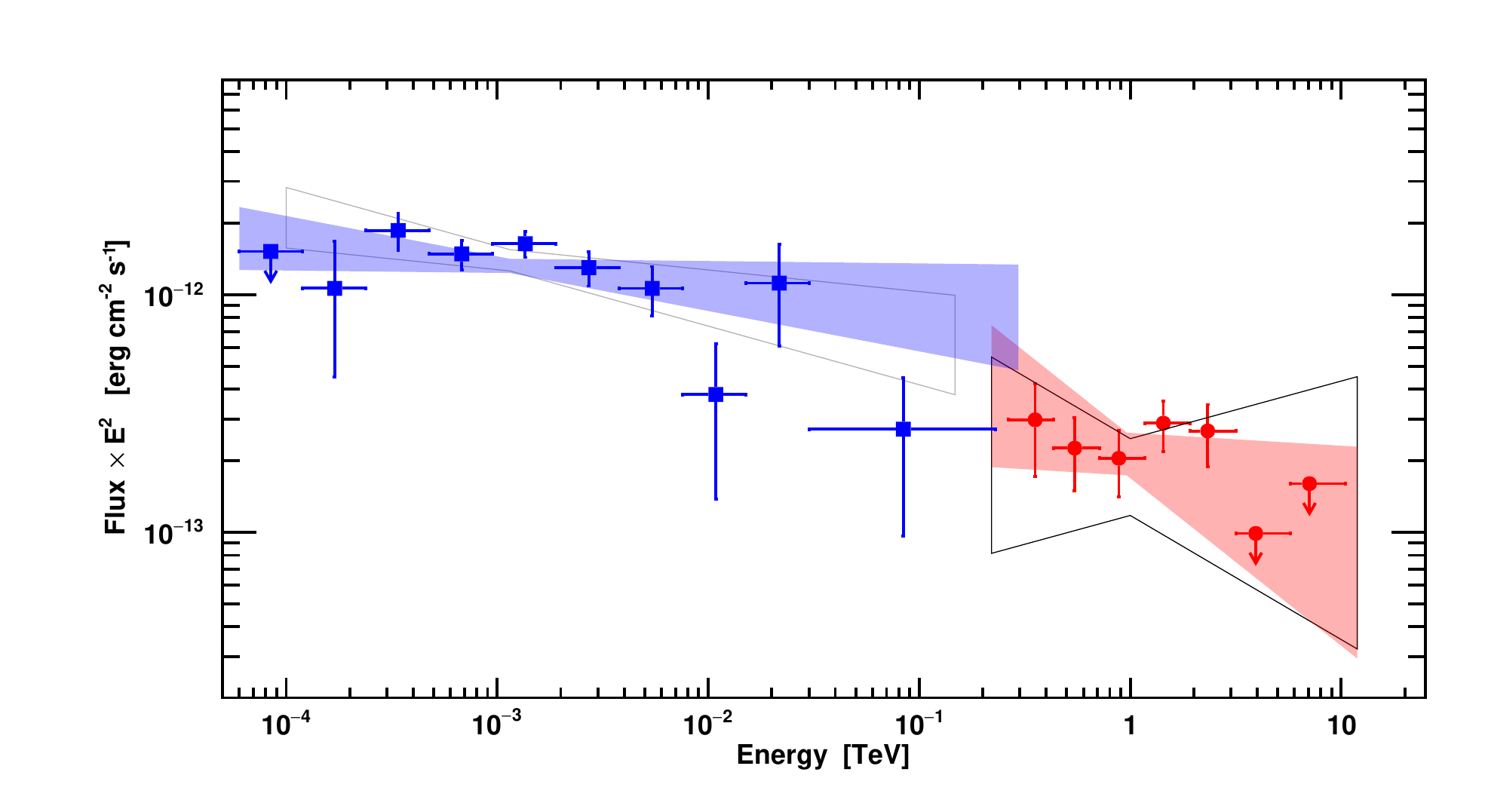}}
\caption{H.E.S.S. and \fermi pass 8 \g-ray SEDs are shown in red and
  blue, respectively. All error bars represent 1$\sigma$ statistical
  uncertainties. The upper limits are given at 95\% confidence
  level. The red shaded area represents the 1$\sigma$ confidence
  region of the H.E.S.S. fit with combined statistical and systematic
  uncertainties. The black box shows the 1$\sigma$ confidence region
  of the H.E.S.S. fit from HESS12. The grey box shows the \fermi 3FGL
  best fit.  We note that the \fermi measurement uncertainties are
  dominated by the low statistics. The systematic error of the \fermi
  points range from 5\,\% to 20\,\%. The blue area shows the best fit
  power-law to the \fermi pass 8 data.}
\label{fig:combinedFit}
\end{center}
\end{figure*}

All uncertainties obtained for the flux normalisation and spectral
index are summarised in Table~\ref{tab:sys_results}. The error bars
for the H.E.S.S. flux points shown in Figure~\ref{fig:combinedFit}
only represent the statistical uncertainties. In this figure, the red
shaded area indicates the combined statistical and systematic
uncertainties of the best-fit power-law model. The black contour
depicts the error region as derived in HESS12.

\begin{table}[htbp]
\caption{Estimated systematic uncertainties of the H.E.S.S. observations towards NGC\,253.}
\begin{center}
\begin{tabular}{@{}lcc@{}}
  \Xhline{3\arrayrulewidth}
  Origin of uncertainty 							& spectral index 	& normalisation \\ \hline
  reconstruction, 								& \multirow{2}{*}{$\pm\,0.19$} & \multirow{2}{*}{2\,\%} \\ 
  calibration \& analysis							&				 &			\\
  run selection									& $\pm\,0.07$		& 10\,\%		 \\
  selection cuts									& $\pm\,0.11$		& 13\,\%		 \\
  atmospheric modelling							& $\pm\,0.09$		& 10\,\%		 \\
  \cline{2-3}
  {\bf Total systematic uncertainty}					& {\bf $\pm$\,0.25}	& \bf{ 19\,\%}		 \\
  \Xhline{3\arrayrulewidth}
\end{tabular}
\end{center}
\label{tab:sys_results}
\end{table}

\section{Fermi-LAT data analysis}
\label{sec:fermi}

Since the publication of HESS12, the dataset accumulated by \fermi has
increased by a factor of two. In addition, the release of the pass 8
data products~\citep{W.Atwood:2013aa} allows for an additional gain in
sensitivity and performance, especially at the lower end of the \fermi
energy range of around 100 MeV. As this is the energy region where
differences between hadronic and leptonic emission processes are best
visible, a new analysis of more than 8 years' \fermi data was
performed.

\subsection{\fermi data}
\fermi observations towards NGC\,253 were selected in the time period
of MET 239557417 - MET 507108410 (corresponding to 04 Aug 2008 - 26
Jan 2017), more than 8 years in total and double the data that was
used in HESS12. The standard {\it Fermi} Science
Tools\footnote{v10r0p5, \url{http://fermi.gsfc.nasa.gov/ssc}} were
employed to carry out the data analysis. A square region-of-interest
(ROI) of $15^{\circ} \times 15^{\circ}$ was defined around the optical
center of NGC\,253. In order to suppress albedo background events from
the Earth's limb, events arriving during times in which the ROI was
observed under unfavourable zenith angles were not included in the
analysis. Specifically, times in which the spacecraft was tilted more
than $52^{\circ}$ and in which the ROI was only observable at zenith
angles $> 90^{\circ}$ were neglected. To avoid biases due to energy-dispersion
effects at low energies, a pre-selection of \g\ rays was
performed that restricts the energy range to a minimum of 30\,MeV.

The HE \g-ray light curve of NGC\,253 was monitored with the high
cadence long-term monitoring tool {\it FlaapLUC}~\citep{lenain2017flaapluc}
which did not reveal significant variability.

\subsection{New HE \g-ray spectrum}
The spectral analysis was performed based on the P8\_R2\_v6 version of
the post-launch instrument response functions (IRFs). Both front and
back converted photons were selected. A binned likelihood analysis
using the {\it gtlike} tool in the energy range from 60\,MeV to
500\,GeV was performed. All known sources as described in the 3FGL, as
well as the diffuse galactic background {\it{ iso\_P8R2\_SOURCE\_V6}}
and the isotropic diffuse background \g-ray emission
({\it{gll\_iem\_v06}}) were included in the fit. The flux
normalisations and spectral indices of all contributing sources in the
ROI were left free during the fit. NGC\,253 is detected with a TS
value of 480, corresponding to roughly 22$\sigma$. The flux above
60\,MeV $F(E>60\, {\rm MeV})$ is
$1.56 \pm 0.28^{stat} \pm 0.15^{sys} \times 10^{-8}~\rm cm^{-2}
s^{-1}$. Assuming the spectrum follows a power-law, the best-fit
spectral index is $2.09 \pm 0.07^{stat} \pm 0.05^{sys}$.

To derive the SED, the energy range from 60\,MeV to 30\,GeV was
divided into nine equally spaced $\mathrm{log}_{10}(E)$ energy bins,
while the higher-energy part from 30\,GeV to 300\,GeV was combined
into a single bin due to the low count statistics at these
energies. The fluxes obtained in these bins are shown in
Fig.~\ref{fig:combinedFit} and were obtained with a likelihood fit
carried out in each bin accounting for the energy dispersion.
The spectral points are given in Table~\ref{fig:combinedFit}.
In the first energy bin ranging from 60 to 120\,MeV NGC\,253 is not
detected significantly. Therefore an upper limit at a confidence level
of 95\% is derived. All higher-energy spectral points have
TS values larger than 4, which corresponds to a significance of more
than 2$\sigma$. A photon with an energy of 214\,GeV was detected
within 0.1 degrees from NGC\,253, which limits the highest-energy
bin to 230 GeV. At energies above
$\sim\,3$\,GeV, the \fermi SED is very well described by a power-law
extending into the entire H.E.S.S. energy domain. A power-law fit to
all data points at energies above 3\,GeV is found to yield a spectral
slope of $2.22\,\pm \, 0.06^{\rm stat}$.

\section{Results and Discussion of the combined HE- and VHE- gamma ray
  spectrum}

\label{sec:Results}
\subsection{Cosmic-ray luminosity and propagation in the starburst}

From the combined {\it Fermi}-H.E.S.S. \g-ray observations in the
energy range from 0.1\,GeV to 3\,TeV, the inferred integrated \g-ray
luminosity is estimated to be
$L_{\gamma}=1.19 \pm 0.35^{\rm stat} \times 10^{40} {\rm
  ~erg~s}^{-1}$. Adopting a fiducial CR luminosity for the nucleus of
the starburst galaxy region of $L_{\rm CR}=10^{41}{\rm ~erg~s}^{-1}$,
it is immediately apparent that of the order of $\sim$10\% of such a
CR luminosity must be transferred to \g-rays. The motivation for this
fiducial CR luminosity comes from the inferred Milky Way's CR
luminosity, which is estimated to lie within the range
$0.6-3\times 10^{41} {\rm ~erg~s}^{-1}$\citep{drury2012origin}.

Further consideration of the reference CR luminosity value for the
starburst region comes from an estimation of the power fed into the CR
population by SNe in this system,
$L_{\rm CR}=\Theta E_{\rm SN}\nu_{\rm SN}\approx 1.6\times 10^{41}{\rm
  ~erg~s}^{-1}$, where a fixed fraction $\Theta\approx 0.1$ of the
supernova remnant (SNR) kinetic power is fed into CRs, a total SNR kinetic energy
$E_{\rm SN}=10^{51}$~erg is assumed to be released in each event, and
the SNR rate within the starburst region of NGC\,253 is
$\nu_{\rm SN}\approx 0.05$~yr$^{-1}$, motivated from radio, infrared (IR), and
optical observations of NGC\,253 and taken from
\citet{engelbracht1998nuclear,van1994more,ohm2013non} for the distance
of $3.5$\,Mpc \citep[see HESS12]{dalcanton09}. This estimated value is
within the estimated range of $(0.6-3)\times 10^{41} {\rm ~erg~s}^{-1}$
for the CR luminosity of the Milky Way under the same assumptions
\citep{drury2012origin}.

\begin{figure*}[tb]
\begin{center}
\resizebox{0.9 \hsize}{!}{\includegraphics{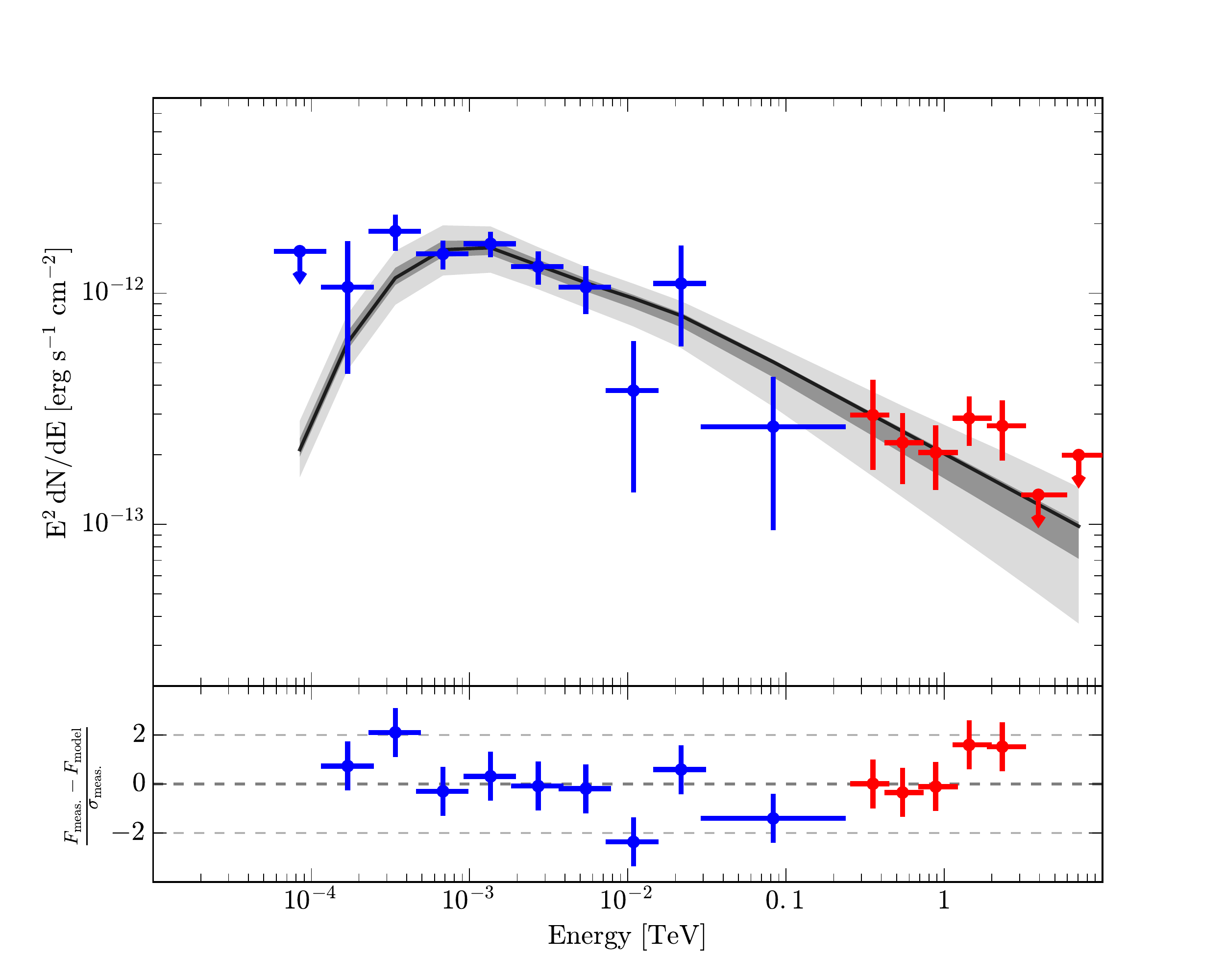}}
\caption{The \g-ray SED obtained with H.E.S.S. and \fermi are shown
  together. In addition, the best-fit \g-ray spectrum from pion decay
  of inelastically scattered protons is shown as a black solid
  line. The grey shaded areas highlight the 1 and 3$\sigma$ confidence
  regions of the fit. The lower panel shows the residuals of the
  measurement with respect to the best fit, normalised to the
  measurements statistical uncertainty. Blue and red data points
  correspond again to the measurements obtained with \fermi and
  H.E.S.S., respectively.}
\label{fig:naima}
\end{center}
\end{figure*}

The rather hard overall differential \g-ray spectrum observed for this
system up to the highest detected energies is an indication against
diffusion-dominated transport of the CRs in the starburst region,
a scenario which would be analogous to the
conventional diffusion picture for diffuse CRs in the ISM of our
Galaxy. Indeed, the high velocity of the starburst wind (see HESS12)
rather suggests an advection-dominated transport, and therefore a
spectrum of \g-rays emitted from the starburst region whose form is
close to that of the source charged-particle spectrum, at least at
energies above a few GeV, where the form of the hadronic \g-ray energy
spectrum should roughly follow that of the generating charged
particles. How closely the form of the \g-ray spectrum follows that of
the CRs depends on the density of the gas in the starburst region. We
consider two scenarios, which represent two extreme cases for this system.\\
\\
If the gas in the system acts as a {\it thick} target, CRs will lose
all their energy in the starburst region through $pp$ interactions. In
this regime, the rise of the $pp$ interaction cross-section with
energy softens the spectrum of CRs in the system relative to their
source spectrum. This softening, however, is naturally compensated for
by the \g-ray emission process itself, resulting in the photon index
matching that of the source CR spectral slope. To ascertain the
best-fit spectral slope for this case, a power-law fit to all data
points from 3\,GeV to 3\,TeV was performed. This approach neglects the
systematic uncertainties of both measurements, resulting in small
statistical uncertainties on the obtained fit parameters. Energies
below 3\,GeV are not considered as the proton kinematics start to
impact the results, leading to a departure from the power-law
description at energies close to the pion production threshold. The
best-fit spectral index was found to be $2.22 \pm 0.06^{\rm stat}$.\\
\\
If the gas in the starburst region is considered to be a {\it thin}
target for CRs, particles are able to escape the starburst region via
advection before loosing a significant fraction of their energy. In this regime, the
spectral slope of CRs in the system is not altered relative to that in
their source. Once again, however, the growth of the $pp$
cross-section with energy results in higher energy CRs more
efficiently losing their energy than lower energy CRs, hardening the
\g-ray spectrum produced.  In turn, correcting for this over-representation
of high energy \g-rays yields a CR spectrum that is
softer than the \g-ray spectrum. In order to estimate the CR spectral
shape under these assumptions, a description of starburst nucleus of
NGC\,253 as well as the inelastic proton scattering cross-section is
necessary. The cross-section and branching ratios can be forward
folded with a proton test distribution. A tool that allows for all of
this is {\it NAIMA}~\citep{zabalza2015naima}, a tool which
employs the Markov Chain Monte Carlo methods
from~\citet{goodman2010ensemble} implemented in {\it
  emcee}~\citep{foreman2013emcee}. It fits the parameters of the
proton test distributions based on the measured \g-ray spectrum,
utilising the \citet{Kafexhiu:2014cua} parameterisation of the $pp$
interaction differential cross-sections. Assuming a distance of
3.5\,Mpc~\citep{dalcanton09} and utilising the cross-section of the
$pp$ energy losses and pion production from PYTHIA~8, protons were
simulated in a kinetic energy range from $0.1$~GeV to $0.5$~PeV
according to a power-law in momentum with spectral index $\alpha$ and
normalisation $N_0$ at the reference momentum $p_0$ of the form
\begin{equation}
  N(E) = \frac{N(p_{0})}{\beta c}\times \left(\frac{p}{p_0}\right)^{-\alpha},
\end{equation}
where $E$ is the total energy of the proton, $p$ is the proton momentum, and
$\beta$ the proton velocity in units of $c$. The fit was done using
both, the updated H.E.S.S. and \fermi spectrum, using statistical
uncertainties only. Also, the upper limits are taken into account in
the fit. The resulting best-fit spectral index obtained is
$\alpha = 2.46 \pm 0.03^{\rm stat}$. The \g-ray spectrum that is
produced via pion decay from this proton distribution is depicted in
Fig.~\ref{fig:naima}. The $\chi^{2}/{\rm N}_{\rm dof}$ of the best
fit of 0.97 for 12 degrees of freedom, as well as the residuals
indicate a good fit to the data. The total energy available in protons
above a kinetic energy of 290\,MeV (the pion production threshold) is
$(2.0 \pm 0.2^{\rm stat}) \times 10^{53}$~erg.

Since the systematic uncertainties on the measured flux normalisation
in the H.E.S.S. energy range are as large as the statistical ones, the
fit was performed again by shifting the H.E.S.S. data points. For
this, all measured flux points were increased (or decreased) by
$20\,\%$. The resulting best fit parameters for the proton spectrum
varied by $\sim \, 3\,\%^{\rm sys}_{\rm H.E.S.S.\,norm}$ and
$\pm 0.03^{\rm sys}_{\rm H.E.S.S.\,norm}$ for the normalisation and
the spectral index, respectively. Additionally, the impact of the
exact choice of {\it pp}-interaction parameterisation was tested by
repeating the fit using alternative descriptions of {\it
  pp}-interaction processes from QGSJET and SYBILL instead of
PYTHIA8. We found that the choice of the parameterisation adds another
uncertainty at the same level as the statistical ones, namely
$\sim \, 6\, \%^{\rm sys}_{\rm interaction}$ and
$\pm 0.03^{\rm sys}_{\rm interaction}$ for the flux normalisation and
the spectral index respectively. If the H.E.S.S. measurement is
excluded from the fit, the best-fit spectral index is around 2.6 and
undershoots the H.E.S.S. measurement.\\
\\
A limiting factor in the effort to recover the CR source spectrum
comes from the complications introduced into the in-situ CR picture,
from a consideration of the potential competing particle propagation
and energy loss channels. The source spectrum will have an index
between those obtained in the {\it thick} and {\it thin} one-dimensional (1D)
models as long as convection dominates the particle propagation in the
starburst region and only pion-decay \g-ray production matters.

\subsection{Cosmic-ray calorimetry in the starburst nucleus}

The level at which the starburst system acts as a CR calorimeter,
$f_{{\rm cal}}$, is defined by the ratio of power that is channelled
into pion production relative to the total amount of potential power
available for pion production. A complete comparison of the
calorimetric level therefore requires the further estimation of the
fraction of CR energy in the population with energies above the
threshold for pion production, $f_{\pi}$. We note that \g-ray
observations are only sensitive to this high-energy component. As was
discussed in HESS12, $f_{\pi}$ is reasonably estimated with
$f_{\pi} \approx 3-\Gamma$~(see Appendix in
HESS12) which is based on the assumption of a simple power-law extrapolation
of the CR momentum spectrum with index $\Gamma$ over the whole
particle energy range.

An estimation of the calorimetric level \mbox{$f_{{\rm cal}} = L_{\pi}/L_{\rm CR}(E_{\rm CR}> E_{\pi}^{\rm th.})$} of the system,
using the reference values and a value $f_{\pi}~\sim 0.66$, intermediate
between the two extreme cases $\Gamma = 2.22$ and $2.46$, gives

\begin{eqnarray}
  f_{\rm cal}\approx 0.34\left(\frac{0.66}{f_{\pi}}\right)
  \left(\frac{L_{\gamma}}{1.2 \times 10^{40} 
  {\rm ~erg~s}^{-1}}\right)\left(\frac{1.6 \times 10^{41} 
  {\rm ~erg~s}^{-1}}{L_{\rm CR}}\right),
\label{cal_val2}
\end{eqnarray}

where $E_{\pi}^{\rm th.}$ is the threshold energy in $pp$ collisions
of CRs with the target gas, and the relation $L_{\pi}=3L_{\gamma}$ has
been assumed. It is important to note, however, that on top of the
uncertainties in the determination of $L_{\gamma}$, considerable
contributions to the uncertainty of this value exist in the
determination of the luminosity of CRs with energies above the pion
production threshold ($f_{\pi}L_{\rm CR}$).

The adoption of the extrema in the estimated range of the values
$L_{\gamma}$, $L_{\rm CR}$, and $f_{\pi}$ in Eq.~\ref{cal_val2} can be
used to estimate the subsequent uncertainty range in the
calorimetric value estimation, $f_{\rm cal}$. These uncertainty
contributions in its derivation collectively broaden the overall
uncertainty, with the corresponding range of values obtained for
NGC\,253 being $f_{{\rm cal}} \approx 0.1 - 1$. As part of this
derivation, we impose a ceiling limit of $1$, since values beyond this
level are considered unphysical. This result highlights that only a
crude order of magnitude estimation of the calorimetric value is
presently possible. As such, this result is compatible with the
estimate of \citet{wang2017starburst} found using older HE and VHE
\g-ray spectra in a more detailed calculation of the {\it thick}
target scenario.

Furthermore, it should be noted that this level of uncertainty prevents a
true estimation of the underlying uncertainty in $L_{\rm CR}$, which
is difficult to estimate in the absence of direct CR (and their
secondaries) observations in external galaxy systems. Our estimation
of the level of uncertainty in this result should therefore be
considered as a lower limit on this range, since these additional
contributions would be expected to further broaden it. Such
considerations highlight the difficulties faced in constraining the
calorimetric value for starburst galaxies, and the essential role
played by high-quality spectral measurements.


\section{Conclusions}

The observational and analysis results presented here strengthen the
interpretation of the \g-ray emission in a hadronic scenario, as
previously considered in \citet{abramowski2012spectral}. Key
supporting findings are based on the improved H.E.S.S. and \fermi
analysis. The deeper understanding of the systematics at energies
above 100\,GeV help to better constrain the spectral shape. The \fermi
pass 8 analysis and a factor two more statistics provide a more
accurate measurement of the \g-ray emission in the energy range below
100\,GeV.

The assumption that a population of protons is giving rise to the
measured \g-ray spectrum through hadronic collisions provides an
excellent fit to the data. Based on the presented analyses, the CR
luminosity in the starburst nucleus was evaluated in two extreme
scenarios assuming the gas to act as either a thin or thick target
for the CRs. These two scenarios allowed us to bracket the CR
luminosity in NGC\,253 with uncertainties of one order of
magnitude. The calorimetric level of NGC\,253 was calculated to lie in
the transition between the two scenarios with allowed values ranging
from $f_{\rm cal} \approx 0.1$ to $1$.

The presented spectra will remain the most precise measurements for
the coming years. The VHE \g-ray spectrum will only be updated once
the Cherenkov Telescope Array (CTA) has started operations and
collected a sizeable set of observations towards NGC\,253. The CTA
measurement will provide more precise and detailed \g-ray data,
potentially yielding an estimate of the extension of the region
emitting VHE \g-rays.
As demonstrated by~\citet{Consortium:2017aa} (their figure 11.4),
the current gap in high-quality \g-ray data from 50\,GeV to 200\,GeV
will be filled down to 100 GeV within 100 hours
of observations. At higher energies, CTA will be able to probe the presence or
absence of a cut-off of the \g-ray spectrum and therefore able to probe the
acceleration limit of the astrophysical particle accelerators in the
starburst nucleus or the onset of \g-\g\,\,absorption in the dense
radiation fields. Further significant improvements in the HE \g-ray
domain will only be possible with missions like e-ASTROGAM, which
could provide an accurate measurement in the energy range below
1\,GeV~\citep{de2017astrogam} and probe the existence of a 'pion-bump'
as predicted by hadronic emission models.


\begin{acknowledgements}
The support of the Namibian authorities and of the University of Namibia in facilitating the construction and operation of H.E.S.S. is gratefully acknowledged, as is the support by the German Ministry for Education and Research (BMBF), the Max Planck Society, the German Research Foundation (DFG), the Alexander von Humboldt Foundation, the Deutsche Forschungsgemeinschaft, the French Ministry for Research, the CNRS-IN2P3, the U.K. Science and Technology Facilities Council (STFC), the Knut and Alice Wallenberg Foundation the National Science Centre, Poland grant no. 2016/22/M/ST9/00382, the South African Department of Science and Technology and National Research Foundation, the University of Namibia, the National Commission on Research, Science \& Technology of Namibia (NCRST),  the Innsbruck University, the Austrian Science Fund (FWF), and the Austrian Federal Ministry for Science, Research and Economy, the University of Adelaide and the Australian Research Council, the Japan Society for the Promotion of Science and by the University of Amsterdam. We appreciate the excellent work of the technical support staff in Berlin, Heidelberg, Palaiseau, Paris, Saclay, and in Namibia in the construction and operation of the equipment. This work benefited from services provided by the H.E.S.S. Virtual Organisation, supported by the national resource providers of the EGI Federation.
\end{acknowledgements}


\bibliographystyle{aa}
\bibliography{ngc_253_bib}

\begin{thebibliography}{40}
\expandafter\ifx\csname natexlab\endcsname\relax\def\natexlab#1{#1}\fi

\bibitem[{{Aartsen} {et~al.}(2014){Aartsen}, {Ackermann}, {Adams}, {Aguilar},
  {Ahlers}, {Ahrens}, {Altmann}, {Anderson}, {Arguelles}, {Arlen}, \&
  et~al.}]{Aartsen2014}
{Aartsen}, M.~G., {Ackermann}, M., {Adams}, J., {et~al.} 2014, Physical Review
  Letters, 113, 101101

\bibitem[{Abramowski {et~al.}(2012)Abramowski, Acero, Aharonian, Akhperjanian,
  Anton, Balzer, Barnacka, Becherini, Becker, Bernl{\"o}hr,
  {et~al.}}]{abramowski2012spectral}
Abramowski, A., Acero, F., Aharonian, F., {et~al.} 2012, The Astrophysical
  Journal, 757, 158

\bibitem[{Acero {et~al.}(2009)Acero, Aharonian, Akhperjanian, Anton,
  De~Almeida, Bazer-Bachi, Becherini, Behera, Bernl{\"o}hr, Bochow,
  {et~al.}}]{acero2009detection}
Acero, F., Aharonian, F., Akhperjanian, A., {et~al.} 2009, Science, 326, 1080

\bibitem[{Ackermann {et~al.}(2012)Ackermann, Ajello, Allafort, Baldini, Ballet,
  Bastieri, Bechtol, Bellazzini, Berenji, Bloom, {et~al.}}]{ackermann2012gev}
Ackermann, M., Ajello, M., Allafort, A., {et~al.} 2012, The Astrophysical
  Journal, 755, 164

\bibitem[{{Aharonian} {et~al.}(2006){Aharonian}, {Akhperjanian}, {Bazer-Bachi},
  {Beilicke}, {Benbow}, {Berge}, {Bernl{\"o}hr}, {Boisson}, {Bolz}, {Borrel},
  {Braun}, {Breitling}, {Brown}, {B{\"u}hler}, {B{\"u}sching}, {Carrigan},
  {Chadwick}, {Chounet}, {Cornils}, {Costamante}, {Degrange}, {Dickinson},
  {Djannati-Ata{\"\i}}, {O'C.~Drury}, {Dubus}, {Egberts}, {Emmanoulopoulos},
  {Espigat}, {Feinstein}, {Ferrero}, {Fiasson}, {Fontaine}, {Funk}, {Funk},
  {Gallant}, {Giebels}, {Glicenstein}, {Goret}, {Hadjichristidis}, {Hauser},
  {Hauser}, {Heinzelmann}, {Henri}, {Hermann}, {Hinton}, {Hofmann}, {Holleran},
  {Horns}, {Jacholkowska}, {de Jager}, {Kh{\'e}lifi}, {Komin}, {Konopelko},
  {Kosack}, {Latham}, {Le Gallou}, {Lemi{\`e}re}, {Lemoine-Goumard}, {Lohse},
  {Martin}, {Martineau-Huynh}, {Marcowith}, {Masterson}, {McComb}, {de
  Naurois}, {Nedbal}, {Nolan}, {Noutsos}, {Orford}, {Osborne}, {Ouchrif},
  {Panter}, {Pelletier}, {Pita}, {P{\"u}hlhofer}, {Punch}, {Raubenheimer},
  {Raue}, {Rayner}, {Reimer}, {Reimer}, {Ripken}, {Rob}, {Rolland}, {Rowell},
  {Sahakian}, {Saug{\'e}}, {Schlenker}, {Schlickeiser}, {Schwanke}, {Sol},
  {Spangler}, {Spanier}, {Steenkamp}, {Stegmann}, {Superina}, {Tavernet},
  {Terrier}, {Th{\'e}oret}, {Tluczykont}, {van Eldik}, {Vasileiadis}, {Venter},
  {Vincent}, {V{\"o}lk}, {Wagner}, \& {Ward}}]{crabpaper}
{Aharonian}, F., {Akhperjanian}, A.~G., {Bazer-Bachi}, A.~R., {et~al.} 2006,
  aap, 457, 899

\bibitem[{Atwood {et~al.}(2013)Atwood, Albert, Baldini, Tinivella, Bregeon,
  Pesce-Rollins, Sgr{\`o}, Bruel, Charles, Drlica-Wagner, Franckowiak, Jogler,
  Rochester, Usher, Wood, Cohen-Tanugi, \& for~the
  Fermi-LAT~Collaboration}]{W.Atwood:2013aa}
Atwood, W., Albert, A., Baldini, L., {et~al.} 2013, arXiv preprint
  arXiv:1303.3514

\bibitem[{{Booth} {et~al.}(2013){Booth}, {Agertz}, {Kravtsov}, \&
  {Gnedin}}]{Booth2013}
{Booth}, C.~M., {Agertz}, O., {Kravtsov}, A.~V., \& {Gnedin}, N.~Y. 2013,
  \apjl, 777, L16

\bibitem[{Breitschwerdt {et~al.}(1991)Breitschwerdt, McKenzie, \&
  V{\"o}lk}]{breitschwerdt1991galactic}
Breitschwerdt, D., McKenzie, J., \& V{\"o}lk, H. 1991, Astronomy and
  Astrophysics, 245, 79

\bibitem[{Breitschwerdt {et~al.}(1993)Breitschwerdt, McKenzie, \&
  V{\"o}lk}]{breitschwerdt1993galactic}
Breitschwerdt, D., McKenzie, J., \& V{\"o}lk, H. 1993, Astronomy and
  Astrophysics, 269, 54

\bibitem[{{CTA Consortium}(2017)}]{Consortium:2017aa}
{CTA Consortium}. 2017, Science with the Cherenkov Telescope Array,
  arxiv:1709.07997

\bibitem[{Dalcanton {et~al.}(2009)Dalcanton, Williams, Seth,
  {et~al.}}]{dalcanton09}
Dalcanton, J.~J., Williams, B.~F., Seth, {et~al.} 2009, ApJS, 183, 67

\bibitem[{De~Angelis {et~al.}(2017)De~Angelis, Tatischeff, Tavani, Oberlack,
  Grenier, Hanlon, Walter, Argan, von Ballmoos, Bulgarelli,
  {et~al.}}]{de2017astrogam}
De~Angelis, A., Tatischeff, V., Tavani, M., {et~al.} 2017, Experimental
  Astronomy, 44, 25

\bibitem[{de~Naurois \& Rolland(2009)}]{de2009high}
de~Naurois, M. \& Rolland, L. 2009, Astroparticle Physics, 32, 231

\bibitem[{Drury(2012)}]{drury2012origin}
Drury, L.~O. 2012, Astroparticle Physics, 39, 52

\bibitem[{{Eichmann} \& {Becker Tjus}(2016)}]{Eichmann2016}
{Eichmann}, B. \& {Becker Tjus}, J. 2016, \apj, 821, 87

\bibitem[{Engelbracht {et~al.}(1998)Engelbracht, Rieke, Rieke, Kelly, \&
  Achtermann}]{engelbracht1998nuclear}
Engelbracht, C., Rieke, M.~J., Rieke, G.~H., Kelly, D., \& Achtermann, J. 1998,
  The Astrophysical Journal, 505, 639

\bibitem[{{Fermi-LAT Collaboration} {et~al.}(2010){Fermi-LAT Collaboration},
  {Abdo}, {Ackermann}, {Ajello}, \& et~al.}]{abdo2010}
{Fermi-LAT Collaboration}, {Abdo}, A.~A., {Ackermann}, M., {Ajello}, M., \&
  et~al. 2010, \apjl, 709, L152

\bibitem[{Foreman-Mackey {et~al.}(2013)Foreman-Mackey, Hogg, Lang, \&
  Goodman}]{foreman2013emcee}
Foreman-Mackey, D., Hogg, D.~W., Lang, D., \& Goodman, J. 2013, Publications of
  the Astronomical Society of the Pacific, 125, 306

\bibitem[{Goodman \& Weare(2010)}]{goodman2010ensemble}
Goodman, J. \& Weare, J. 2010, Communications in applied mathematics and
  computational science, 5, 65

\bibitem[{{H.E.S.S. Collaboration}(2015)}]{LMC2015}
{H.E.S.S. Collaboration}. 2015, Science, 347, 406

\bibitem[{Indriolo \& McCall(2013)}]{Indriolo2013}
Indriolo, N. \& McCall, B.~J. 2013, Chem. Soc. Rev., 42, 7763

\bibitem[{Kafexhiu {et~al.}(2014)Kafexhiu, Aharonian, Taylor, \&
  Vila}]{Kafexhiu:2014cua}
Kafexhiu, E., Aharonian, F., Taylor, A.~M., \& Vila, G.~S. 2014, Phys. Rev.,
  D90, 123014

\bibitem[{{Kennicutt} \& {Evans}(2012)}]{Kennicutt2012}
{Kennicutt}, R.~C. \& {Evans}, N.~J. 2012, \araa, 50, 531

\bibitem[{Krumholz(2014)}]{Krumholz2014}
Krumholz, M.~R. 2014, Physics Reports, 539, 49

\bibitem[{Lenain(2017)}]{lenain2017flaapluc}
Lenain, J.-P. 2017, Astrophysics Source Code Library

\bibitem[{Ohm \& Hinton(2013)}]{ohm2013non}
Ohm, S. \& Hinton, J. 2013, Monthly Notices of the Royal Astronomical Society:
  Letters, 429, L70

\bibitem[{{Pakmor} {et~al.}(2016){Pakmor}, {Pfrommer}, {Simpson}, \&
  {Springel}}]{Pakmor2016}
{Pakmor}, R., {Pfrommer}, C., {Simpson}, C.~M., \& {Springel}, V. 2016, \apjl,
  824, L30

\bibitem[{{Papadopoulos} \& {Thi}(2013)}]{Papadopoulos2013}
{Papadopoulos}, P.~P. \& {Thi}, W.-F. 2013, in Advances in Solid State Physics,
  Vol.~34, Cosmic Rays in Star-Forming Environments, ed. D.~F. {Torres} \&
  O.~{Reimer}, 41

\bibitem[{Parker(1966)}]{parker1966dynamical}
Parker, E. 1966, The Astrophysical Journal, 145, 811

\bibitem[{{Persic} {et~al.}(2008){Persic}, {Rephaeli}, \&
  {Arieli}}]{Persic2008}
{Persic}, M., {Rephaeli}, Y., \& {Arieli}, Y. 2008, \aap, 486, 143

\bibitem[{{Rephaeli} {et~al.}(2010){Rephaeli}, {Arieli}, \&
  {Persic}}]{Rephaeli2010}
{Rephaeli}, Y., {Arieli}, Y., \& {Persic}, M. 2010, \mnras, 401, 473

\bibitem[{{Salem} \& {Bryan}(2014)}]{Salem2014a}
{Salem}, M. \& {Bryan}, G.~L. 2014, \mnras, 437, 3312

\bibitem[{Salem {et~al.}(2016)Salem, Bryan, \& Corlies}]{salem2016role}
Salem, M., Bryan, G.~L., \& Corlies, L. 2016, Monthly Notices of the Royal
  Astronomical Society, 456, 582

\bibitem[{{Torres} {et~al.}(2012){Torres}, {Cillis}, {Lacki}, \&
  {Rephaeli}}]{Torres2012}
{Torres}, D.~F., {Cillis}, A., {Lacki}, B., \& {Rephaeli}, Y. 2012, \mnras,
  423, 822

\bibitem[{Van~Buren \& Greenhouse(1994)}]{van1994more}
Van~Buren, D. \& Greenhouse, M.~A. 1994, The Astrophysical Journal, 431, 640

\bibitem[{{VERITAS Collaboration} {et~al.}(2009){VERITAS Collaboration},
  {Acciari}, {Aliu}, {Arlen}, \& et~al.}]{acciari2009}
{VERITAS Collaboration}, {Acciari}, V.~A., {Aliu}, E., {Arlen}, T., \& et~al.
  2009, \nat, 462, 770

\bibitem[{Wang \& Fields(2017)}]{wang2017starburst}
Wang, X. \& Fields, B.~D. 2017, Monthly Notices of the Royal Astronomical
  Society, 474, 4073

\bibitem[{{Wik} {et~al.}(2014){Wik}, {Lehmer}, {Hornschemeier}, {Yukita},
  {Ptak}, {Zezas}, {Antoniou}, {Argo}, {Bechtol}, {Boggs}, {Christensen},
  {Craig}, {Hailey}, {Harrison}, {Krivonos}, {Maccarone}, {Stern}, {Venters},
  \& {Zhang}}]{Wik2014}
{Wik}, D.~R., {Lehmer}, B.~D., {Hornschemeier}, A.~E., {et~al.} 2014, \apj,
  797, 79

\bibitem[{Zabalza(2015)}]{zabalza2015naima}
Zabalza, V. 2015, arXiv preprint arXiv:1509.03319

\bibitem[{Zirakashvili {et~al.}(1996)Zirakashvili, Breitschwerdt, Ptuskin, \&
  V{\"o}lk}]{zirakashvili1996magnetohydrodynamic}
Zirakashvili, V., Breitschwerdt, D., Ptuskin, V., \& V{\"o}lk, H. 1996,
  Astronomy and Astrophysics, 311, 113

\end{thebibliography}

\begin{table}
\begin{tabular}{ll}
\hline \hline
Min. - Mean - Max. & Energy Flux \\
 Energy [$\mathrm{TeV}$] &  [$\mathrm{erg\,s^{-1}\,cm^{-2}}$] \\
\hline \hline
$(6.00 $ - $8.45 $ - $12.0) \times 10^{-5}$ & $\,< 1.52 \times 10^{-12}$  \\
$(1.20 $ - $1.70$ - $2.39) \times 10^{-4}$  & $(1.06  \pm 0.62) \times 10^{-12}$ \\
$(2.39$ - $3.39 $ - $4.76) \times 10^{-4}$  & $(1.86  \pm 0.33) \times 10^{-12}$ \\
$(4.76 $ - $6.79 $ - $9.50) \times 10^{-4}$ & $(1.48  \pm 0.21) \times 10^{-12}$ \\
$(0.95$ - $1.36 $ - $1.89) \times 10^{-3}$  & $(1.64  \pm 0.20) \times 10^{-12}$ \\
$(1.89$ - $2.72 $ - $3.78) \times 10^{-3}$  & $(1.30  \pm 0.21) \times 10^{-12}$ \\
$(3.78 $ - $5.43 $ - $7.54) \times 10^{-3}$ & $(1.06  \pm 0.25) \times 10^{-12}$ \\
$(0.75 $ - $1.09 $ - $1.50) \times 10^{-2}$ & $(3.79  \pm 2.42) \times 10^{-13}$ \\
$(1.50$ - $2.17 $ - $3.00) \times 10^{-2}$  & $(1.10  \pm 0.51) \times 10^{-12}$ \\
$~0.03 $ - $0.08$ - $0.23$ & $(2.65  \pm 1.71) \times 10^{-13}$ \\ \hline
$~0.26$ - $0.35$ - $0.43$ & $(2.97  \pm 1.25) \times 10^{-13}$ \\
$~0.43$ - $0.55$ - $0.71$ & $(2.26  \pm 0.77) \times 10^{-13}$ \\
$~0.71$ - $0.88$ - $1.17$ & $(2.05  \pm 0.64) \times 10^{-13}$ \\
$~1.17$ - $1.43$ - $1.93$ & $(2.88  \pm 0.70) \times 10^{-13}$ \\
$~1.93$ - $2.33$ - $3.17$ & $(2.67  \pm 0.78) \times 10^{-13}$ \\
$~3.17$ - $3.93$ - $5.75$ & $\,< 1.34 \times 10^{-13}$\\
$~5.75$ - $7.07$ - $10.4$ & $\,< 1.99 \times 10^{-13}$\\
\hline
\hline
\end{tabular}
\caption{The $\gamma$-ray spectral data as displayed in
Figure~\ref{fig:combinedFit} and~\ref{fig:naima}. The {\it Fermi} and
H.E.S.S. spectral points are separated by the horizontal line. The energy
flux uncertainties correspond to the statistical uncertainties only.
The flux limits are calculated for a confidence interval of 95\,\%.}
\label{tab:spec_points}
\end{table}

\end{document}